\documentclass[pra,twocolumn,showpacs]{revtex4}
\usepackage{amsmath,amssymb,bm,txfonts,amssymb,epsfig,graphics}
\usepackage{CJK}

\begin{document}



\title{Coupling mechanism between microscopic two-level system and
superconducting qubits}
\author{Zhen-Tao Zhang}
\author{Yang Yu}
\affiliation{National Laboratory of Solid State Microstructures, School of Physics,
Nanjing University , Nanjing 210093 , China}
\date{\today}
\pacs{03.67.Lx 85.25.Cp 03.65.Yz}

\begin{abstract}
We propose a scheme to clarify the coupling nature between superconducting
Josephson qubits and microscopic two-level systems. Although dominant
interests of studying two-level systems were in phase qubits previously, we
find that the sensitivity of the generally used spectral method in phase
qubits is not sufficient to evaluate the exact form of the coupling. On the
contrary, our numerical calculation shows that the coupling strength changes
remarkably with the flux bias for a flux qubit, providing a useful tool to
investigate the coupling mechanism between the two-level systems and qubits.
\end{abstract}

\maketitle







Recent progress on superconducting qubits suggested that superconducting
Josephson circuits are promising candidates for practical quantum
computating \cite{Makhlin01,Clarke08,Neeley10,DiCarlo10,Sun10a}. However,
extensive works are needed to understand the decoherence mechanism hence
increase the decoherence time of these macroscopic quantum systems. For
instance, microscopic defects are ubiquitous in solid state devices. Each of
these defects may behavior empirically as a quantum two-level-system (TLS)
with characteristic frequency ranging about several gigahertz. The
anticrossings resulting from the resonance between the TLSs and qubit were
observed in the spectra of phase qubit \cite{Simmonds04,Sun10b}, flux qubit
\cite{Plourde05}, and charge qubit \cite{Schreier08}. Experiments have shown
that TLSs not only shorten the decoherence time \cite{Cooper04}, but also
reduce the visibility of Rabi oscillation of the qubit, limiting the
fedelity of the quantum gate. Moreover, an ensemble of TLSs with various
characteristic frequencies may produce low frequency $1/f$ noise \cite%
{Shnirman05}. Therefore, it is imperative to understand the microscopic
coupling mechanism between TLS and superconducting qubits.

Unfortunately, it is nearly unattainable to directly probe a single TLS's
microscopic mechanism due to its microscopic nature. Nevertheless, utilizing
qubit as a detector of TLS supplied an alternative method to extract useful
information \cite{Cooper04,Martinis05,Yu08}. Based on the experimental
results, three coupling models between TLS and superconducting qubit were
suggested: critical current fluctuator \cite{Simmonds04}, electric dipole
\cite{Martinis05}, and flux fluctuator \cite{Cole10}. Great effort has been
put to determine which is the exact coupling mechanism \cite%
{Martinis05,Tian07,Kim08,Lupascedilcu09,Cole10}. Recently, it is suggested
that one may clarify the coupling mechanism by investigating the
longitudinal component of the interaction Hamiltonian \cite%
{Lupascedilcu09,Cole10}. Lupa\c{s}cu \emph{et al}. have measured the flux
qubit at symmetric double-well potential and only found transverse term \cite%
{Lupascedilcu09}. Furthermore, Cole \emph{et al}. have theoretically shown
that for different models, the coupling form between qubit and TLS changes
remarkably. For the electric dipole model the coupling type is totally
transverse, while for the other two models, besides the transverse, a weak
longitudinal coupling exists \cite{Cole10}. However, their experiments in a
flux-biased phase qubit is unable to determine the coupling type, because
the observed longitudinal coupling strength is too small to discriminate it
from the experimental uncertainty. Therefore, two questions come out from
their works. One is whether we have to consider both transverse and
longitudinal terms simultaneously for all superconducting qubits coupled
with TLS. The other is whether the magnitude of the two terms depends on the
coupling model thereby we can determine the exact coupling model by
measuring the two terms. In this work, we have analyzed the interaction
Hamiltonian of the three models. We found that the transverse and
longitudinal terms vary for different types of superconducting qubits. For
phase qubits the longitudinal coupling is always small thereby it is not a
good system to probe the coupling form. However, the longitudinal term for
flux qubits is comparable to the transverse term. In addition, for different
coupling models the longitudinal term exhibits distinct flux bias
dependencies, supplying a hopeful scheme to clarify the microscopic model of
the TLS.\newline
\indent We start from a short review of the three models which are used to
describe the microscopic nature of the coupling between TLS and qubits.
Although they are also valid for flux qubits, we at first discuss them in a
flux biased phase qubit for simplicity. The first model is critical current
fluctuator. In this model, the microscopic TLS was assumed as a critical
current fluctuator whose two states respond to two different critical
currents of the Josephson junction \cite{Simmonds04}. The fluctuator could
be considered as a ion moving between the left and right well in a double
well potential. If $\hbar \varepsilon $ is the energy difference between the
two position states, $\Delta $ is the tunneling matrix element, the
interaction Hamiltonian can be written as \cite{Cole10}
\begin{equation*}
H_{I}=\upsilon _{i}\cos \hat{\phi}(\cos \theta \sigma _{T}^{x}+\sin \theta
\sigma _{T}^{z})
\end{equation*}%
with $\sigma _{T}^{x,z}$ being Pauli operators of TLS in its eigenenergy
space. $\upsilon _{i}=-\frac{\delta I_{0}\phi _{0}}{2\pi },$ where $\delta
I_{0}$ represents the difference of the critical currents for the ion
populating the right and left well, $\phi_0$ is flux quantization. $\hat{\phi}$ is phase difference across
the Josephson junction. $\theta $ denotes the relative orientation of the
TLS's configuration basis and eigenbasis, $\tan \theta =\varepsilon /\Delta $%
. Although in many literatures it is assumed $\theta =0$ \cite%
{Simmonds04,Ku05}, we would like to keeping the above formula which is more
accurate and general.\newline
\indent The second model is the interaction of an electrical dipole with
field \cite{Martinis05}. TLS has been modeled as an electric charge
distribution that changed between the two states. In Ref. \cite{Martinis05},
each TLS was assumed as an electron of charge e at position R or L separated
by a distance $d,$ resulting a dipole moment $\mu =ed$. The interaction
Hamiltonian is given by
\begin{equation*}
H_{I}=\overrightarrow{\mu }\cdot \overrightarrow{E}=\upsilon _{q}\hat{q}%
(\cos \theta \sigma _{T}^{x}+\sin \theta \sigma _{T}^{z})\cos \eta
\end{equation*}%
where $\upsilon _{q}=\frac{2e^{2}d}{Cx}$. $x$ is the thickness of the
junction. $\hat{q}$ represent the number of Cooper-pairs tunneled across the
junction. $\eta $ denote the relative angle between electric dipole $%
\overrightarrow{\mu }$ and the electric field $\overrightarrow{E}$. The
defination of $\theta $ is terminologically the same as that in the last
model. However, $\varepsilon $ and $\Delta $ may depend on different
physical variables. It is worth to note that the coupling term in this model
is purely transverse no matter whether the double well of the TLS is
symmetric \cite{Martinis05,Lupascedilcu09,Cole10}. This is a useful
characteristic which offers a possibility to discriminate this coupling
mechanism from others.\newline
\indent Another reasonable model for TLS is magnetic flux fluctuator. It has
been found nearly three decades ago that the flux embraced by a rf-SQUID
fluctuates with frequency lying in low frequency regime. The flux
fluctuation is responsible for the dephasing of superconducting qubits.
Recently, several experiments were carried out to probe the flux noise and
found that the spectral density of the low frequency flux noise with the
form $1/f$ \cite{Yoshihara06,Kakuyanagi07,Bialczak07}. Moreover, Shnirman
\emph{et al}. have shown that a collective of TLS's with a natural
distribution accounts for both high frequency and low frequency $1/f$ noise
\cite{Shnirman05}. Therefore, a new explanation for TLS in superconducting
phase qubit was proposed, suggesting that the states of the TLS may modulate
the magnetic flux $\phi _{e}$ threading the superconducting loop \cite%
{Cole10}. The resulted coupling is on the variable $\hat{\phi}$ for the
phase qubit. The interaction Hamiltonian can be written as%
\begin{equation*}
H_{I}=\upsilon _{\phi }\hat{\phi}(\cos \theta \sigma _{T}^{x}+\sin \theta
\sigma _{T}^{z})
\end{equation*}%
where $\upsilon _{\phi }=-\frac{\delta \phi _{e}}{L}(\frac{\phi _{0}}{2\pi }%
)^{2}$. $\delta \phi _{e}$ is the difference of $\phi _{e}$ between the two
configuration states of TLS. $\theta $ is defined the same as before, with $%
\varepsilon $ and $\Delta $ relying on different physical variables.\newline
\begin{figure}[tbp]
\includegraphics [width=7.5cm,height=6cm]{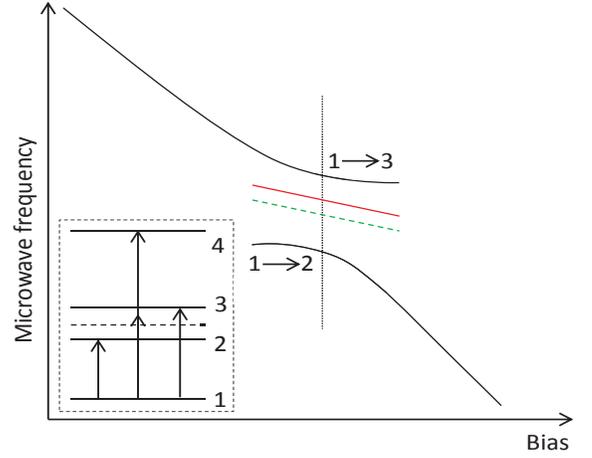}
\caption{(Color online) Schematic diagram of the spectrum of a phase qubit
resonantly coupled to a TLS. The green dashed line and red thin solid line
represent the resonant peaks of the two-photon transition from state
1 to state 4 of the composite system without or with longitudinal
coupling, respectively. Without longitudinal coupling, the
two-photon spectrum locates at the exact middle of $1\rightarrow2$ and $1\rightarrow3$ one-photon transition
lines. Note that with longitudinal coupling the two-photon spectrum can move up or
down relative to the middle position, corresponding to positive or negetive sign of
the longitudinal coupling. Here we chose positive sign as an example.
Inset: the energy levels of the resonant qubit-TLS system.}
\end{figure}
\indent The experimental investigation on the coupling mechanism is not very
convincing. Recently, Cole $et$ $al.$ proposed a scheme to probe the
coupling mechanism in a phase qubit \cite{Cole10}. Therein, whether the
longitudinal coupling exists is a crucial clue to decide which model is
true. The longitudinal coupling between resonant qubit-TLS leads to asymmetry of the
two-photon transition spectrum relative to the one-photon transitions
spectra: $\omega _{1\leftrightarrow 4}\neq \omega _{1\leftrightarrow
2}+\omega _{1\leftrightarrow 3}$ (1-4 denote the eigenstates of the coupled
TLS-qubit system), as shown in Fig. 1. Therefore, one could experimentally
resolve the coupling type of qubit-TLS system via spectral analysis.
However, their experiment can not confirm the existence of the longitudinal
coupling because its value is comparable to the measurement uncertainty.
Hence, they could not reach a conclusion on the correct model. In our
opinion, the key reason for the ambiguity is that the longitudinal is much
smaller than the transverse coupling in all three models. To demonstrate
this, we unify the interaction Hamiltonian
\begin{equation*}
H_{I}=\upsilon _{k}\hat{O}(\cos \theta \sigma _{T}^{x}+\sin \theta \sigma
_{T}^{z})
\end{equation*}%
where $k=i,\phi ,q$ and $\hat{O}=\cos \hat{\phi},\hat{\phi},\hat{q}$. Ignore
the mixed terms such as $\sigma _{q}^{z}\sigma _{T}^{x}$, which have no
effect on the spectrum of the system, we can write the interaction
Hamiltonian in the eigenenergy basis
\begin{equation*}
H_{I}=\upsilon _{k}(o_{x}^{k}\cos \theta \sigma _{q}^{x}\sigma
_{T}^{x}+o_{z}^{k}\sin \theta \sigma _{q}^{z}\sigma _{T}^{z})
\end{equation*}%
where the factors $o_{x,z}^{k}$ are given by
\begin{equation}
o_{x}^{k}=\frac{|\langle 1|\hat{O}|0\rangle +\langle 0|\hat{O}|1\rangle |}{2}%
,\indent o_{z}^{k}=\frac{|\langle 1|\hat{O}|1\rangle -\langle 0|\hat{O}%
|0\rangle |}{2}  \notag
\end{equation}%
For the electric dipole model, $\hat{q}$ has no diagonal elements, $%
o_{z}^{q}=0$, so there is no longitudinal coupling in this case. Turning to the other two models, we
numerically calculated $o_{x,z}^{i,\phi}$ as functions of flux bias using
the parameters in Ref. \cite{Cole10}, shown in Fig. 2. It is found that $\frac{o_{z}^{i,\phi }}{o_{x}^{i,\phi }}<\frac{1}{6}$. Moreover, for $\phi_e>0.6$, the qubit can not work
due to the large tunneling rate of the excite state; for $\phi_e<0.57$,
$o_{z}^{i,\phi}$ is at least one order of magnitude smaller than $o_{x}^{i,\phi}$.The much
smaller longitudinal coupling factor relative to the transverse one may be the main reason for which one can not
verify the existence of the longitudinal interaction between qubit and TLS.
Even worse, the corresponding coupling factors of the two models are roughly equal. This
is easy to understand if we notice that in phase qubit, $\phi \simeq \frac{%
\pi }{2}$. We substitute $\phi $ with $\frac{\pi }{2}+\varphi $, where $%
\varphi $ is a small quantity,
\begin{equation*}
\cos \phi =\cos (\frac{\pi }{2}+\varphi )=-\sin \varphi \simeq -\varphi =%
\frac{\pi }{2}-\phi
\end{equation*}%
\begin{figure}[tbp]
\includegraphics [width=8.5cm]{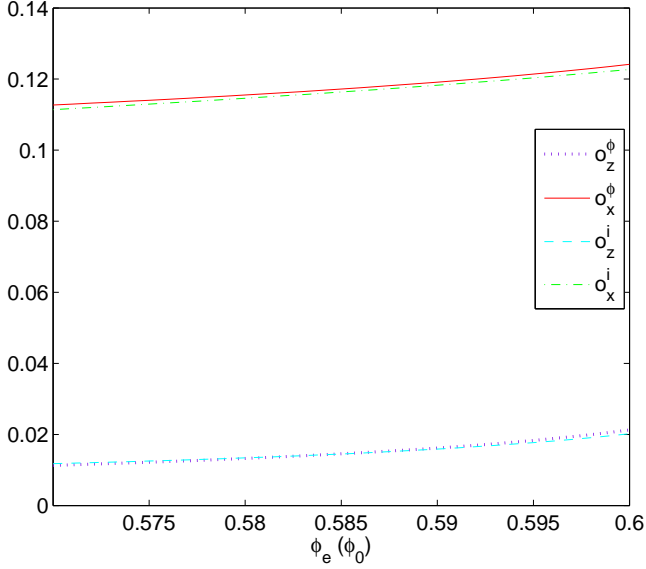}
\caption{(Coler Online) Coupling factors of qubit- TLS interaction in phase
qubitb for the critical current fluctuator and flux fluctuator models. The
parameters of the phase qubit are: $C=850fF,L=720pH,I_{0}=984nA$. The green
dash-dotted and cyan dashed line denote the transverse and longitudinal
coupling factor $o_{x}^{i},o_{z}^{i}$ of the critical current fluctuator
model respectively. The red solid and purple dotted line denote $o_{x}^{%
\protect\phi },o_{z}^{\protect\phi }$ of the flux fluctuator model
respectively.}
\end{figure}
Taking $\cos \phi \simeq \frac{\pi }{2}-\phi $ into the coupling factors
expressions, we can obtain: $o_{x}^{i}\simeq o_{x}^{\phi }$, $%
o_{z}^{i}\simeq o_{z}^{\phi }$. Therefore, for phase qubits, the
longitudinal coupling is not sensitive to the coupling mechanism. Then, it is
difficult to clarify the coupling nature between TLS and phase qubit .
Although one may argue that the TLS parameter $\theta $ has a crucial effect
on the longitudinal coupling magnitude, unfortunately, till now, people are unable to control the angle $\theta $ due to the poor knowledge of TLS.


\indent Instead of phase qubit, we find that flux qubit is a possible system to
reveal the coupling nature of TLS and qubits. We start from a flux qubit
which consists of a superconducting loop interrupted by three Josephson
junctions \cite{Orlando99}. Two junctions are the same and the other one is $%
\alpha $ times smaller than them. If we assume that the large junctions have
a critical current $I_{0}$ and a capacitance $C$, then the critical current
and capacitance of the small junction are $\alpha I_{0}$ and $\alpha C,$
respectively. The Hamiltonian of the circuit is \cite{Orlando99}
\begin{eqnarray}
H_{q} &=&4E_{c}{\hat{n}_{1}}^{2}-E_{J}\cos {\hat{\phi}_{1}}+4E_{c}{\hat{n}_{2}}^{2}-E_{J}\cos
{\hat{\phi} _{2}}+\frac{4E_{c}}{\alpha }{\hat{n}_{3}}^{2}  \notag \\
&&-\alpha E_{J}\cos {\hat{\phi} _{3}}+E_{J}(2+\alpha )
\end{eqnarray}%
where $E_{c}=\frac{e^{2}}{2C}$, $E_{J}=I_{0}\phi _{0}/2\pi $. $\hat{\phi}
_{i} \;(i=1,2,3)$ is the phase difference across each junction, and its
conjugate variable $\hat{n}_{i}$ is the number of Cooper pair through each
junction. If the external magnetic flux $\phi_{ext}=f\phi_{0}$, using the
flux quantization condition, we get $\hat{\phi}_{3}=2\pi f+\hat{\phi}_{1}-\hat{\phi} _{2}$.
Transforming the coordinates $\hat{\phi} _{1}$, $\hat{\phi} _{2}$ to the sum and the
difference of the phases, $\hat{\phi} _{p}=(\hat{\phi}_{1}+\hat{\phi}_{2})/2,\,\hat{\phi}
_{m}=(\hat{\phi} _{1}-\hat{\phi} _{2})/2$, $H_{q}$ is reduced to
\begin{eqnarray}
H_{q} &=&E_{p}{\hat{n}_{p}}^{2}+E_{m}{\hat{n}_{m}}^{2}-2E_{J}\cos {\hat{\phi} _{p}}\cos {\hat{\phi}
_{m}}+E_{J}(2+\alpha)   \notag \\
&&-\alpha E_{J}\cos (2\pi f+2\hat{\phi} _{m}))\nonumber
\end{eqnarray}%
where $E_{p}=2E_{c},\,\hat{n}_{p}=-i\frac{\partial }{\partial \phi _{p}}$, $%
E_{m}=E_{p}/(1+2\alpha ),\,\hat{n}_{m}=-i\frac{\partial }{\partial \phi _{m}}$. We
have calculated the lowest three energy levels near $f=0.5$ , with typically
chosen parameters $E_{J}/E_{c}=40$, $\alpha =0.68$. In the region $0.49<f<0.51
$, the energy difference between the lowest two levels (qubit) is much
smaller than that between the upper two levels, showing a very good
nonlinearity which enables the spectroscopic experiment will not involve the
third level of the qubit. \newline
\indent In the three-junction flux qubit, each junction has the possibility
of containing TLS. Even though, we can prove numerically that the location
of the TLS in different junctions would not affect our results
qualitatively. Therefore, we consider that the TLS is in the small junction
without losing generality. For electric dipole model, the interaction term
is
\begin{eqnarray}
H_{I} &=&\vec{\mu}\cdot \vec{E}  \notag \\
&=&\frac{de}{x}\hat{V}(\cos \theta \sigma _{T}^{x}+\sin \theta \sigma
_{T}^{z})\cos \eta  \notag \\
&=&\frac{de}{x}\frac{\phi _{0}}{2\pi }2\dot{\hat{\phi}}_{m}(\cos \theta \sigma
_{T}^{x}+\sin \theta \sigma _{T}^{z})\cos \eta  \notag
\end{eqnarray}%
Using Heisenberg equation $\dot{\hat{\phi}}_m=\frac{1}{i\hbar }[\hat{\phi} _{m},H_{q}]
$, we obtain
\begin{equation}\label{4}
H_{I}=\frac{\hbar \omega _{q}d\langle 0|\hat{\phi} _{m}|1\rangle }{x}\cos \eta
\cos \theta \cdot \sigma _{q}^{x}\sigma _{T}^{x}
\end{equation}%
where $\hbar \omega _{q}$ is the eigenenergy of the flux qubit. Obviously,
similar to that in phase qubit the longitudinal coupling is zero.

For the other models, following the same procedures used in the phase qubit,
we can straightforwardly write out the interaction Hamiltonians.\newline
Critical current model:
\begin{eqnarray}
H_{I} &=&-\frac{\alpha \phi _{0}\delta I_{0}}{2\pi }\cos (2\pi f+2\hat{\phi}
_{m})(\cos \theta \sigma _{T}^{x}+\sin \theta \sigma _{T}^{z})  \notag \\
&=&-\frac{\alpha \phi _{0}\delta I_{0}}{2\pi }(o_{x}^{i}\cos \theta \sigma
_{q}^{x}\sigma _{T}^{x}+o_{z}^{i}\sin \theta \sigma _{q}^{z}\sigma _{T}^{z})
\end{eqnarray}%
\begin{eqnarray}
o_{x}^{i} &=&\frac{|\langle 1|\cos (2\pi f+2\hat{\phi} _{m})|0\rangle +\langle
0|\cos (2\pi f+2\hat{\phi} _{m})|1\rangle |}{2}  \notag \\
o_{z}^{i} &=&\frac{|\langle 1|\cos (2\pi f+2\hat{\phi} _{m})|1\rangle -\langle
0|\cos (2\pi f+2\hat{\phi} _{m})|0\rangle |}{2}  \notag
\end{eqnarray}%
Magnetic flux fluctuator model
\begin{eqnarray}
H_{I} &=&2\pi \alpha E_{J}\delta\phi_e\sin (2\pi f+2\hat{\phi} _{m})(\cos \theta \sigma
_{T}^{x}+\sin \theta \sigma _{T}^{z})  \notag \\
&=&2\pi \alpha E_{J}\delta\phi_e(o_{x}^{\phi }\cos \theta \sigma _{q}^{x}\sigma
_{T}^{x}+o_{z}^{\phi }\sin \theta \sigma _{q}^{z}\sigma _{T}^{z})
\end{eqnarray}%
\begin{eqnarray}
o_{x}^{\phi } &=&\frac{|\langle 1|\sin (2\pi f+2\hat{\phi} _{m})|0\rangle +\langle
0|\sin (2\pi f+2\hat{\phi} _{m})|1\rangle |}{2}  \notag \\
o_{z}^{\phi } &=&\frac{|\langle 1|\sin (2\pi f+2\hat{\phi} _{m})|1\rangle -\langle
0|\sin (2\pi f+2\hat{\phi} _{m})|0\rangle |}{2}  \notag
\end{eqnarray}%
\begin{figure}[tbp]
\includegraphics [width=8cm]{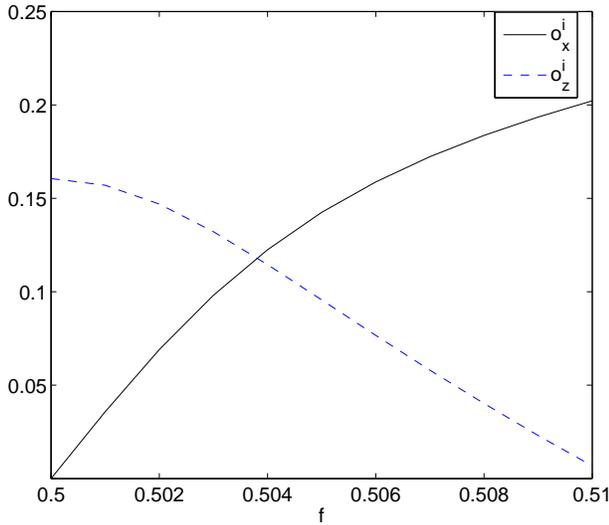}
\caption{(Color online) Transverse (solid line) and longitudinal (dashed
line) coupling factors for critical current fluctuator model.}
\end{figure}
\begin{figure}[tbp]
\includegraphics [width=8cm]{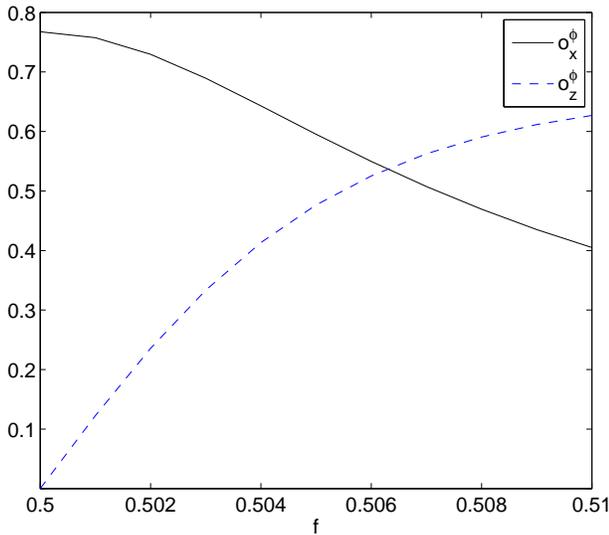}
\caption{(Color online) Transverse (solid line) and longitudinal (dashed
line) coupling factors for flux fluctuator model.}
\end{figure}
\indent Now we compare the magnitudes of the transverse and longitudinal
couplings. As discussed before, currently it is impossible to change the
orientation of the TLS basis, we focus on the factors $o_{x,z}^{i,\phi }$.
Using the same parameters as above ($E_{J}/E_{c}=40$, $\alpha =0.68$), we
have numerically calculated $o_{x,z}^{i,\phi }$ as functions of $f,$ shown
in Fig. 3 and 4. When $f$ varies from 0.5 to 0.51, $o_{x,z}^{i,\phi }$
exhibit remarkable changes with totally different trends. For critical
current fluctuator model (Fig. 3), the transverse coupling factor $o_{x}^{i}$
is zero at the degenerate point $f=0.5$ while the longitudinal factor
reaches the maximum. Then, away from $f=0.5$, the transverse factor
increases monotonically while the longitudinal one decreases gradually. At $%
f=0.51$, the transverse factor becomes much larger than the longitudinal
one. In recent experiments \cite{Lupascedilcu09}, the splitting resulted
from transverse coupling is observed at the degeneracy point of a flux
qubit. In addition, no longitudinal coupling is observed at $f=0.5$. These
behaviors disagree with the predictions of the critical current fluctuator
model, indicating that the critical current fluctuation is not the dominate
mechanism of the qubit-TLS coupling.\newline
\indent For the flux fluctuator model, the trends are totally converse (Fig.
4). The transverse factor reaches maximum at the degenerate point while the
longitudinal magnitude vanishes, indicating that the coupling is pure
transverse at this point. However, the electric dipole model predicts
similar pure transverse coupling [see Eq. (\ref{4})]. we can not
discriminate the flux fluctuator and the electric dipole model at the
degenerate point. When we tune the flux bias away from $f=0.5$, the
transverse factor decreases and the longitudinal one increases gradually. At
$f=0.51$, the longitudinal factor is larger than $1.5$ times of the
transverse one. Therefore, the flux fluctuator model contains both
transverse and longitudinal coupling but in the electric dipole model only
transverse interaction exists. We can clarify the microscopic mechanism of
TLS by studying the coupling term of TLS-flux qubit interaction in a flux
qubit biased away from the degenerate point. In practical, TLSs have been
observed in three-junction flux qubits biased away from the degenerate point
\cite{Plourde05}, suggesting that this spectral method is completely
feasible with the current technique.\newline
\indent In summary, we have calculated the qubit-TLS coupling factors of
both transverse and longitudinal terms under three microscopic models. It is
found that for phase qubits the longitudinal coupling is difficult to
observe because it is always much smaller than the transverse coupling.
Then, we show that in three-junction flux qubit the relative magnitude of
the transverse and longitudinal coupling factors are largely model-dependent
and very sensitive to the external flux bias. We propose that these features
can be used to clarify the microscopic model of TLS.

This work was supported in part by MOST (2011CB922104, 2011CBA00200), NSFC
(91021003, 10725415), and the Natural Science Foundation of Jiangsu Province
(BK2010012).


\end{document}